\documentclass[fleqn,twoside]{article}
\usepackage{espcrc2}
\usepackage{graphicx}
\usepackage[figuresright]{rotating}
\newcommand{\AmS}{{\protect\the\textfont2
  A\kern-.1667em\lower.5ex\hbox{M}\kern-.125emS}}

\newcommand{\ba}{\begin{array}{c}}
\newcommand{\baz}{\begin{array}{cc}}
\newcommand{\bad}{\begin{array}{ccc}}
\newcommand{\bav}{\begin{array}{cccc}}
\newcommand{\bea}{\begin{equation} \begin{array}{c}}
\newcommand{\eea}{ \end{array} \end{equation}}
\newcommand{\ea}{\end{array}}

\newcommand{\be}{\begin{eqnarray*}}
\newcommand{\ee}{\end{eqnarray*}}
\newcommand{\beq}{\begin{equation}}
\newcommand{\eeq}{\end{equation}}
\newcommand{\dm}{\mbox{$\Delta m_{21}^2$~}}
\newcommand{\dmsol}{\mbox{$\Delta m^2_{\odot}$~}}

\newcommand{\kl}{\mbox{KL~}}

\newcommand{\sss}{\sin^2 \theta_{12}}
\newcommand{\dma}{\mbox{$\Delta m^2_{\rm A}$}}
\newcommand{\meff}{\mbox{$\left| < \! m \! > \right|$}}
\newcommand{\mefff}{\mbox{$\langle m \rangle$}}
\newcommand{\betabeta}{\mbox{$(\beta \beta)_{0 \nu}$}}
\newcommand{\hbeta}{$\mbox{}^3 {\rm H}$ $\beta$-decay }

\def\gs{\mathrel{
   \rlap{\raise 0.511ex \hbox{$>$}}{\lower 0.511ex \hbox{$\sim$}}}}
\def\ls{\mathrel{
   \rlap{\raise 0.511ex \hbox{$<$}}{\lower 0.511ex \hbox{$\sim$}}}}
\def\ltap{\ \raisebox{-.4ex}{\rlap{$\sim$}} \raisebox{.4ex}{$<$}\ }
\def\gtap{\ \raisebox{-.4ex}{\rlap{$\sim$}} \raisebox{.4ex}{$>$}\ }

\newcommand{\pmns}{\mbox{$ U_{\rm PMNS}$}}

\title{Towards Complete Neutrino Mixing Matrix and CP-Violation}
 
\author{S. T. Petcov
\address{Scuola Internazionale Superiore di Studi Avanzati, and
         INFN, Trieste, Italy}
        \thanks{Also at: Institute of Nuclear Research and Nuclear Energy,
            Bulgarian Academy of Sciences, 1784 Sofia, Bulgaria}~
        \thanks{Plenary talk given at $\nu$'04 International 
Conference, June 14-19, 2004, Paris, France.}
}

\vspace{-0.8cm}
\begin{document}
\begin{abstract}
 The compelling experimental evidences for 
oscillations of solar, atmospheric and 
reactor neutrinos imply the existence of 
3-neutrino mixing in vacuum.
We review the phenomenology 
of 3-neutrino mixing, and the 
current data on the 3-neutrino mixing
parameters. The opened questions
and the main goals of future 
research in the field of neutrino 
mixing and oscillations are outlined.
A phenomenological approach 
for understanding the pattern 
of neutrino mixing 
as an interplay between
the mixing, arising from 
the charged lepton sector, and
bimaximal mixing, arising from
a neutrino Majorana mass matrix,
is considered with emphasis 
on the $CP-$violating case.
We comment also on planned future 
steps in the experimental studies of
$\nu$-mixing.

\end{abstract}

\maketitle

\vspace{-0.6cm}
\section{Introduction}

\vspace{-0.2cm}
  The hypothesis of neutrino oscillations 
was formulated in \cite{BPont57}.
In \cite{BPont67} it was suggested that 
the solar $\nu_e$ 
can take part in 
oscillations involving another active
or sterile neutrino. The evidences of 
solar neutrino ($\nu_{\odot}$-) oscillations 
obtained first in the Homestake 
experiment and strengthened by 
the results of Kamioaknde, 
SAGE and GALLEX/GNO 
experiments \cite{sol,Kam96},
were made compelling 
in the last several years by the data
of  Super-Kamiokande (SK), 
SNO and KamLAND (KL) experiments
\cite{SKsol02,SNO123,KL162}.
Under the plausible 
assumption of CPT-invariance, 
the results of the KL reactor
neutrino experiment \cite{KL162} 
established the large mixing 
angle (LMA) MSW 
oscillations/transitions \cite{MSW}
as the dominant mechanism
at the origin of the observed 
solar $\nu_e$ deficit.
The Kamiokande experiment \cite{Kam96}
provided  the first evidences
for oscillations of atmospheric 
$\nu_{\mu}$ and $\bar{\nu}_{\mu}$,
while the data of the 
Super-Kamiokande experiment
made the case of atmospheric 
neutrino oscillations convincing 
\cite{SKatmnu04}.
Evidences for oscillations 
of neutrinos were obtained also 
in the first long baseline
accelerator neutrino experiment 
K2K \cite{K2K}. Indications 
for $\nu$-oscillations
were reported by the LSND 
collaboration \cite{LSND}.

 The recent new SK data on the 
$L/E$-dependence of 
multi-GeV $\mu$-like  atmospheric 
neutrino events \cite{SKatmnu04},
$L$ and $E$ being the distance traveled 
by neutrinos and the $\nu$ energy,
and the new 
spectrum data of KL and K2K experiments 
\cite{KL766,K2Knu04},
presented at this Conference,
are the latest significant 
contributions to the 
remarkable progress 
made in the last several
years in the studies of 
$\nu$-oscillations.
For the first time the data
exhibit directly the effects of the 
oscillatory dependence on $L/E$ and $ E$ of 
the probabilities of  
$\nu$-oscillations in vacuum \cite{BP69}.
We begin to ``see''the 
oscillations of neutrinos.
As a result of these
magnificent developments, 
the oscillations of solar $\nu_e$,
atmospheric $\nu_{\mu}$ and
$\bar{\nu}_{\mu}$, 
accelerator $\nu_{\mu}$ (at $L\sim$250 km)
and reactor $\bar{\nu}_e$ (at $L\sim$180 km), 
driven by nonzero $\nu$-masses 
and $\nu$-mixing, can be considered as 
practically established.

\vspace{-0.35cm}
\section{The Neutrino Mixing Parameters}

\vspace{-0.10cm}
   The SK atmospheric neutrino 
and K2K data are
best described in 
terms of dominant 2-neutrino
$\nu_{\mu} \rightarrow \nu_{\tau}$ 
($\bar{\nu}_{\mu} \rightarrow \bar{\nu}_{\tau}$)
vacuum oscillations.
The best fit values and the 
99.73\% C.L. allowed ranges of the 
atmospheric neutrino 
($\nu_{\rm A}$-)
oscillation parameters
read \cite{SKatmnu04}: \\
$|\dma|$=2.1$\times10^{-3}~{\rm eV^2}$,~
$\sin^22\theta_{\rm A}$ = 1.0, \\ 
$|\dma|$=(1.3 -- 4.2)$\times 10^{-3}~$eV$^2$,
$\sin^22\theta_{\rm A} \geq$ 0.85.\\
\noindent The sign of $\dma$ and of 
$\cos2\theta_{\rm A}$, if
$\sin^22\theta_{\rm A} \neq 1.0$, 
cannot be determined using
the existing data. 
The latter implies that when, e.g., 
$\sin^22\theta_{\rm A} = 0.92$, 
one has $\sin^2\theta_{\rm A}\cong 0.64~{\rm or}~ 0.36$.

  The combined 2-neutrino oscillation 
analysis of the solar neutrino 
and the new KL 766.3 Ty spectrum  
data shows \cite{KL766,BCGPRKL2} that the 
$\nu_{\odot}$-oscillation parameters 
lie in the low-LMA region :\\  

\vspace{-0.3cm}
\noindent $\dmsol$=($7.9^{+0.6}_{-0.5}$)$\times 10^{-5}~{\rm eV^2}$, 
$\tan^2 \theta_\odot$=$(0.40^{+0.09}_{-0.07})$.\\

\vspace{-0.3cm}
\noindent The value of $\dmsol$ is 
determined with a remarkably 
high precision. 
The high-LMA solution is excluded at 
more than 3$\sigma$. 
Maximal $\nu_{\odot}$-mixing
is ruled out at $\sim 6\sigma$;
at 95\% C.L., 
$\cos 2\theta_\odot \geq 0.27$. 
One also has: $\dmsol/|\dma| \sim 0.04 \ll 1$. 
  
   The interpretation of the solar and
atmospheric neutrino, and of K2K and KL
data in terms of $\nu$-oscillations requires
the existence of 3-$\nu$ mixing
in the weak charged lepton current:\\

\vspace{-0.2cm}
$\nu_{l \mathrm{L}}  = \sum_{j=1}^{3} U_{l j} \, \nu_{j \mathrm{L}},~~
l  = e,\mu,\tau$,\\

\vspace{-0.2cm}
\noindent where $\nu_{j \mathrm{L}}$ is the 
field of neutrino $\nu_j$ having 
a mass $m_j$ and
$U$ is the Pontecorvo-Maki-Nakagawa-Sakata (PMNS) 
$\nu$-mixing matrix \cite{BPont57,MNS62}.
All existing $\nu$-oscillation data, except
the data of LSND experiment  
\footnote{In the
LSND experiment indications 
for oscillations
$\bar \nu_{\mu}\to\bar \nu_{e}$  
with $(\Delta m^{2})_{\rm{LSND}}\simeq 
1~\rm{eV}^{2}$ were obtained. 
The  LSND results are being tested 
in the MiniBooNE experiment
\cite{MiniB}.}
\cite{LSND},
can be described assuming 
3-$\nu$ mixing in vacuum
and we will consider only 
this possibility. 
The minimal 4-$\nu$ mixing scheme 
which could incorporate the 
LSND indications for 
$\nu$-oscillations
is strongly disfavored by the data \cite{Maltoni4nu}.
The $\nu$-oscillation explanation of the LSND results
is possible assuming 5-$\nu$ mixing \cite{JConrad}.

 The PMNS matrix  can be 
parametrized by 3 angles 
and, depending on whether the massive 
neutrinos $\nu_j$ are 
Dirac or Majorana particles,
by 1 or 3 CP-violation ($CPV$) phases 
\cite{BHP80,SchValle80D81}.
 In the standardly used parameterization 
(see, e.g., \cite{BPP1}),\\
%

\vspace{-0.2cm}
$\pmns = V(\theta_{12},\theta_{13},\theta_{23},\delta)
~{\rm diag}(1, e^{i \alpha}, e^{i \beta})$,\\

\vspace{-0.2cm}
\noindent where 
$V(\theta_{12},\theta_{13},\theta_{23},\delta)$
is a CKM-like matrix, the angles
$\theta_{ij} = [0,\pi/2]$,
$\delta = [0,2\pi]$ is the 
Dirac $CPV$ phase and
$\alpha,\beta$ are two Majorana 
$CPV$ phases \cite{BHP80,SchValle80D81}. 
One can identify 
$\dmsol = \Delta m^2_{21} > 0$.
In this case 
$|\dma|$=$|\Delta m^2_{31}|\cong |\Delta m^2_{32}|$,
$\theta_{12} = \theta_{\odot}$, 
$\theta_{23} = \theta_{\rm A}$.
The angle $\theta_{13}$ is limited by 
the data from the CHOOZ and Palo Verde
experiments~\cite{CHOOZPV}. The limit
depends strongly on $|\dma|$
(see, e.g, \cite{SNO3BCGPR}).
The existing $\nu_{\rm A}$-data
is essentially insensitive 
to $\theta_{13}$ obeying the 
CHOOZ limit \cite{SKatmnu04}.
The probabilities of survival of 
reactor $\bar{\nu}_e$ and solar $\nu_e$,
relevant for the interpretation of
the KL, CHOOZ and $\nu_{\odot}$- data, depend on
$\theta_{13}$:\\

\vspace{-0.2cm}
\noindent $P^{3\nu}_{\rm KL} \cong \sin^4\theta_{13} + 
\cos^4\theta_{13}\left [1 - \sin^2 2\theta_{12}
\sin^2\frac{\Delta m^2_{21}{\rm L}}{4E} \right ]$,\\
$P^{3\nu}_{\rm CHOOZ} \cong 1 -  
\sin^2 2\theta_{13}
\sin^2\frac{\Delta m^2_{\rm 31}{\rm L}}{\rm 4E}$,\\

\vspace{-0.2cm}
$P^{3\nu}_{\odot} \cong \sin^4\theta_{13} + 
\cos^4\theta_{13}~P^{2\nu}_{\odot}
(\Delta m^2_{21},\theta_{12};\theta_{13})$,\\

\vspace{-0.2cm}
\noindent where $P_{\odot}^{2\nu}$ 
is the 2-$\nu$ mixing solar $\nu_e$
survival probability \cite{SP88,SPJRich89,PKSP88,SP97,NAOsc99} 
in the case of transitions driven by
$\Delta m^2_{21}$ and $\theta_{12}$,
in which the solar $e^-$ number density $N_e$
is replaced by $N_e \cos^2\theta_{13}$ \cite{3nuSP88},\\

\vspace{-0.3cm}
$P^{2\nu}_{\odot} = 
 \bar{P}^{2\nu}_{\odot} + P^{2\nu}_{\odot~{\rm osc}}$,
 
\vspace{-0.3cm}
\begin{equation}
\ba
\bar{P}^{2\nu}_{\odot} = \frac {1}{2} + 
 (\frac{1}{2} - P')\cos2\theta_{12}^m(t_0)\cos 2\theta_{12}, \\ [0.3cm]
P'= {{\exp(-2\pi r_{0}{\Delta m^2_{21}\over {2E}}\sin^2\theta_{12})
          - \exp(-2\pi r_{0}{\Delta m^2_{21}\over {2E}})}
         \over {1 - \exp(-2\pi r_{0}{\Delta m^2_{21}\over {2E}})}}
\ea
\label{Psolexp}
\end{equation}
%
\vspace{-0.2cm}
\noindent Here $\bar{P}^{2\nu}_{\odot}$
is the average probability \cite{Parke86,SP88,SP97},
$P^{2\nu}_{\odot~{\rm osc}}$ is 
an oscillating term \cite{SP88,SPJRich89,SP97,NAOsc99},
$P'$ is the ``double exponential'' 
jump probability \cite{SP88} 
and $r_0$ is the ``running''
scale-height of the change of $N_e$ along the 
$\nu$-trajectory in the Sun
\footnote{The claims in \cite{HWLS04} that
in the LMA region
``The double exponential formula is not valid...
It requires production of neutrinos far above the 
resonance region in the density scale. 
This formula is not applicable in the range 
$\Delta m^2 \cos2\theta/(2E)\sim (1.6-8.0)\times
10^{-6}~{\rm eV^2/MeV}$ for which the density
in the production point turns out to be close
to the resonance density.'' are incorrect and/or
misleading. The analyses and the extensive numerical
studies performed in \cite{PKSP88,HaxtonExp95,NAOsc99}
show that expression (\ref{Psolexp}) for
$\bar{P}^{2\nu}_{\odot}$ provides
a high precision description 
of the average solar $\nu_e$ 
survival probability in the Sun
for any values of $\Delta m^2_{21}$ and $\theta_{12}$
(the relevant error does not exceed 
$\sim$(2-3)\%), including the values from the 
LMA region. 
Actually, it follows from the results 
in \cite{HWLS04} that 
the use of the double 
exponential expression for $P'$ 
\cite{SP88}, eq. (\ref{Psolexp}),
for description of the LMA transitions
brings in an imprecision in 
$\bar{P}^{2\nu}_{\odot}$ which does 
not exceed $\sim 10^{-6}$.
Similarly, the claim in \cite{HWLS04}
that ``... for the LMA solution, 
when the final mixing angle is large, 
one cannot use the Landau-Zener probability 
as an approximation for $P_c$. '', $P_c$ being the
jump probability, is  incorrect,
see \cite{LZ,PKSP88,HaxtonExp95}.
}
\cite{SP88,PKSP88,NAOsc99}.
In the LMA solution region 
one has \cite{SPJRich89}
$P^{2\nu}_{\odot~{\rm osc}} \cong 0$.
Using the 3$\sigma$ allowed range
of $|\dma|$=$|\Delta m^2_{31}|$ \cite{SKatmnu04}
and performing a combined analysis of
the solar neutrino, CHOOZ and KL data, 
one finds \cite{BCGPRKL2}:

\vspace{-0.3cm}
$$\sin^2\theta_{13} < 0.05,~~~~99.73\%~{\rm C.L.}$$ 

\noindent Similar constraint is obtained 
from a global 3-$\nu$ 
\begin{figure}[htb]
\vskip -0.2cm
\includegraphics[width=7.0cm,height=6cm,clip=]{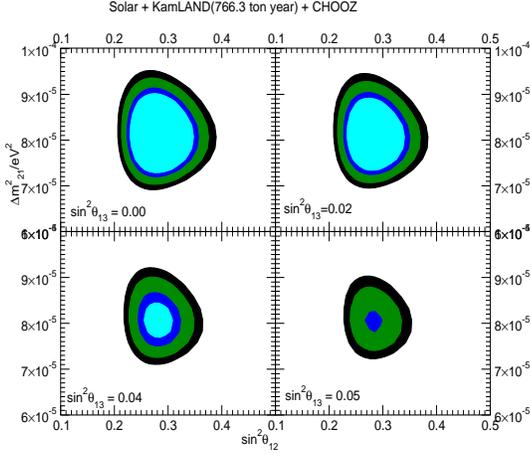}
\vspace{-0.6cm}
\caption{\label{cont3g}
The 90\%, 95\%, 99\% and 99.73\% C.L. allowed 
regions in the $\dm$-$\sss$ plane, obtained 
in a 3-$\nu$ oscillation analysis of the 
solar neutrino, KL and CHOOZ data
\cite{BCGPRKL2}. 
}
\end{figure}

\noindent  oscillation analysis
of the data \cite{Maltoni4nu,3nuGlobal}.
In Fig. (\ref{cont3g})  we show 
the allowed regions in the $\dm-\sss$ plane
for few fixed values of $\sin^2\theta_{13}$ \cite{BCGPRKL2},
obtained using, in  particular, 
eq. (\ref{Psolexp}) for $\bar{P}^{2\nu}_{\odot}$.

  Thus, the fundamental parameters 
characterizing the 3-neutrino mixing are:\\  
$\bullet$ the 3 angles $\theta_{12}$, $\theta_{23}$, $\theta_{13}$,\\
$\bullet$ depending on the nature of $\nu_j$ - 1 Dirac ($\delta$), 
or 1 Dirac + 2 Majorana ($\delta,\alpha,\beta$), 
$CPV$ phases, and\\ 
$\bullet$ the 3 neutrino masses, $m_1,~m_2,~m_3$.

  It is convenient to express the two 
larger masses in terms of the 
third mass and the measured 
$\dmsol$=$\Delta m^2_{21}>$0 
and $\dma$. In the convention 
we are using, the two possible signs of
$\dma$ correspond to two 
types of $\nu$-mass spectrum:\\
$\bullet$ with normal hierarchy,
$m_1 < m_2 < m_3$, $\dma$=$\Delta m^2_{31} >0$,
$m_{2(3)}$=$(m_1^2 + \Delta m^2_{21(31)})^{1\over{2}}$,and\\
$\bullet$ with inverted hierarchy,
$m_3 < m_1 < m_2$,  $\dma$=$\Delta m^2_{32}<$0,
$m_{2}$=$(m_3^2 - \Delta m^2_{32})^{1\over{2}}$, etc.\\
The spectrum can also be \\
$\bullet$ {\it normal hierarchical (NH)}: 
$m_1 \ll m_2 \ll m_3$,\\
\noindent $m_2 \cong (\dmsol)^{1\over{2}}\sim$0.009 eV, 
$m_3 \cong |\dma|^{1\over{2}}\sim$0.045; or\\
$\bullet$ {\it inverted hierarchical (IH)}:
$m_3 \ll m_1 < m_2$, 
with $m_{1,2} \cong |\dma|^{1\over{2}}\sim$0.045 eV; or \\
$\bullet$ {\it quasi-degenerate (QD)}:
$m_1 \cong m_2 \cong m_3 \cong m_0$,
$m_j^2 \gg |\dma|$, $m_0 \gtap 0.20$ eV. 

 After the spectacular experimental 
progress made 
in the studies of $\nu$-oscillations, 
further understanding
of the structure of the $\nu$-masses 
and $\nu$-mixing, of their origins  
and of the status of the CP-symmetry in 
the lepton sector requires a large and
challenging program of research to be pursued 
in neutrino physics.  
The main goals of this 
research program should include:\\
$\bullet$ High precision 
measurement of
the solar and atmospheric  
neutrino oscillations parameters, 
$\Delta m^2_{21}$, $\theta_{21}$, 
and $\Delta m^2_{31}$, $\theta_{23}$.\\
$\bullet$ Measurement of, 
or improving by at least a factor 
of (5 - 10) the existing upper 
limit on, $\theta_{13}$ - 
the only small mixing angle  in 
$\pmns$.\\
$\bullet$ Determination of the
$sign(\Delta m^2_{31})$ and of
the type of $\nu$-mass spectrum 
($NH,IH,QD$, etc.).\\
$\bullet$ Determining or 
obtaining significant constraints
on the absolute scale of $\nu$-masses, 
or on $min(m_j)$.
$\bullet$ Determining 
the nature--Dirac or Majorana,
of massive neutrinos $\nu_j$.\\ 
$\bullet$ Establishing whether the CP-symmetry 
is violated in the lepton 
sector a) due to the Dirac phase $\delta$, and/or
b) due to the Majorana phases 
$\alpha$ and $\beta$ if
$\nu_j$ are Majorana particles.\\
$\bullet$ Searching with increased sensitivity
for possible manifestations, other than 
flavour neutrino oscillations,
of the non-conservation
of the individual lepton charges $L_l$, $l=e,\mu,\tau$,
such as $\mu \rightarrow e + \gamma$,
$\tau \rightarrow \mu + \gamma$, etc. decays.\\
$\bullet$ Understanding at fundamental level 
the mechanism giving rise to  
neutrino masses and mixing and to 
$L_l-$non-conservation,
i.e., finding the {\it Theory of neutrino
mixing}. This includes
understanding the origin of the 
patterns of $\nu$-mixing and $\nu$-masses 
suggested by the data.
Are the observed patterns of 
$\nu$-mixing and of $\Delta m^2_{21,31}$
related to the existence of new  
fundamental symmetry of particle interactions?
Is there any relations between quark 
mixing and neutrino mixing, e.g.,
does the relation $\theta_{12} + \theta_{c}$=$\pi/4$,
where $\theta_{c}$ is the Cabibbo angle, hold?
Is $\theta_{23} = \pi/4$, or $\theta_{23} > \pi/4$
or else $\theta_{23} < \pi/4$?
What is the physical origin 
of $CPV$ phases in $\pmns$?
Is there any relation (correlation)
between the (values of) $CPV$ phases 
and mixing angles in  $\pmns$?
Progress in the theory of $\nu$-mixing
might also lead, in particular, 
to a better understanding of the 
mechanism of generation 
of baryon asymmetry of 
the Universe \cite{LeptoG}.

  Obviously, the successful realization 
of the experimental part of this 
research program would be a formidable task
and would require many years.

The mixing angles, $\theta_{21}$,
$\theta_{23}$ and $\theta_{13}$,
Dirac $CPV$ phase $\delta$ and 
$\Delta m^2_{21}$ and $\Delta m^2_{31}$ 
can, in principle, be measured with a 
sufficiently high precision in
a variety of $\nu$-oscillation 
experiments (see further).
These experiments, however,
cannot provide information on the
absolute scale of $\nu$- masses 
and on the nature of massive neutrinos $\nu_j$.
The flavour neutrino oscillations  
are insensitive to
the Majorana $CPV$ phases 
$\alpha$ and $\beta$ \cite{BHP80,Lang86}.
Establishing whether $\nu_j$
have distinct antiparticles (Dirac fermions)
or not (Majorana fermions) is of fundamental
importance for understanding 
the underlying symmetries of 
particle interactions \cite{BiPet87} 
and the origin of $\nu$-masses. 
If $\nu_j$ are Majorana fermions,
getting experimental information about the  
Majorana $CPV $phases in $\pmns$
is a remarkably challenging problem. 
\cite{BGKP96,MajPhase1,PPR1}. 
The phases $\alpha$ and $\beta$ 
can affect significantly 
the predictions for the 
rates of the (LFV) decays
$\mu \rightarrow e + \gamma$,
$\tau \rightarrow \mu + \gamma$, etc.
in a large class of supersymmetric theories
with see-saw mechanism of neutrino mass
generation (see, e.g., \cite{PPY03}).
Majorana $CPV$ phases might be at 
the origin of the baryon asymmetry of 
the Universe \cite{LeptoG}. 

\vspace{-0.30cm}
\section{The Pattern of Neutrino Mixing}

\vspace{-0.2cm}
 The $\nu$-oscillation data suggest that 
$\theta_{12}\cong \pi/6$, $\theta_{23} \cong \pi/4$, and 
$\theta_{13} \equiv \epsilon < \pi/12$. Thus, the PMNS matrix 
is very different from the  CKM matrix:
\vspace{-0.1cm}
\begin{equation}
\nonumber
\pmns \cong \left(\begin{array}{ccc}
  \frac{\sqrt{3}}{2}& \frac{1}{2} & \epsilon \\
 -\frac{1}{2\sqrt{2}} & \frac{\sqrt{3}}{2\sqrt{2}} & \frac{1}{\sqrt{2}} \\
  \frac{1}{2\sqrt{2}} & -\frac{\sqrt{3}}{2\sqrt{2}} & \frac{1}{\sqrt{2}}
 \end{array} \right)
\end{equation}
%
\noindent where the $CPV$ phases 
have been suppressed.
Understanding the origin of the 
emerged patterns 
of $\nu$-mixing and of 
$\Delta m^2_{21,31}$, 
is one of the central problems 
in today's neutrino physics. 

  $\pmns$ is close in form to 
a bimaximal mixing matrix,
for which $\theta_{12}$=$\theta_{23}$=$\pi/4$ 
and $\theta_{13}$=0:
\vspace{-0.1cm}
\begin{equation}
\nonumber
U_{\rm bimax} =
\left(\begin{array}{ccc}
  \frac{1}{\sqrt{2}} & \frac{1}{\sqrt{2}} & 0 \\
 -\frac{1}{2} & \frac{1}{2} & \frac{1}{\sqrt{2}} \\
  \frac{1}{2} & -\frac{1}{2} & \frac{1}{\sqrt{2}}
 \end{array} \right)
\end{equation}
%
Whereas the data favor $\theta_{23} = \pi/4$
and allows for $\theta_{13} = 0$, 
$\theta_{12} = \pi/4$ is ruled out 
at $\sim$6$\sigma$ \cite{KL766,BCGPRKL2}.
The deviation of $\theta_{12}$ from $\pi/4$
can be parameterized with a 
parameter $\lambda$, which is 
very similar in value 
to the Cabibbo angle \cite{WR}:\\

\vspace{-0.3cm}
$\sin\theta_{21} \cong \frac{1}{\sqrt{2}}( 1 - \lambda),~
\lambda \cong (0.20 - 0.25) 
\sim \theta_C$.\\

\vspace{-0.3cm}
\noindent The implied relation $\theta_\odot = \pi/4 - \theta_C$,
if confirmed experimentally, might be linked to 
GUT's \cite{marttiSMFM}.

It is natural to suppose that \cite{STP82PD,FPR}
\footnote{Extensive list of references on the subject 
is given in \cite{FPR}.} 

\vspace{-0.3cm}
$$\pmns = U^{\dagger}_{\rm L}(\lambda)~U_\nu,~~
U_\nu = U_{\rm bimax},$$

\vspace{-0.1cm}
\noindent where $U^{\dagger}_{\rm L}(\lambda)$ 
arises from diagonalization of 
the charged lepton mass matrix
and $U_{\nu}$ diagonalizes 
a neutrino Majorana mass matrix $M_{\nu}$.
The inequality $\Delta m^2_{21} \ll |\Delta m^2_{31}|$ 
and the form of $U_{\rm bimax}$ can  
be associated with an
{\it approximate} symmetry of $M_{\nu}$, 
implying (for $U_{\rm L}$=${\bf 1}$)
the conservation of the lepton 
charge \cite{STP82PD}:

\vspace{-0.3cm}
$$L' = L_e - L_{\mu} - L_{\tau}~~~~~~~~~~~(4)$$

\vspace{-0.1cm}
  If $\theta^{l}_{ij}$ are the 3 angles of
CKM-like parametrization of $U_{\rm L}(\lambda)$ and
$\sin\theta^{l}_{ij} \equiv \lambda_{ij}$,
3 generic cases are compatible with 
the $\nu$-mixing data \cite{FPR}:\\
$\bullet$ all $\lambda_{ij}$ small, $\lambda_{ij} \ltap 0.35$;\\
$\bullet$$\lambda_{23} = 1$, $\lambda_{12},\lambda_{13} \ltap 0.35$;\\
$\bullet$ all $\lambda_{ij}$ large, e.g., 
$\lambda_{ij} \geq \frac{1}{\sqrt{2}}$.\\
For $\lambda_{ij} \ltap 0.35$, the data imply:
$\lambda_{23} \ltap 0.19$,
$ \lambda_{12} \gg \lambda_{13}$ (for $\lambda_{12,13} > 0$),
$\lambda_{12} \cong $(0.21-0.25), 
$\lambda_{13} \ltap 0.03$.
With all $CPV$ phases set to 0, 
$\theta_{ij}$
are expressed in terms of $\lambda_{kl}$
and their values are correlated.
The deviation of $\theta_{12}$ from $\pi/4$ is 
determined by $s_{13}$=$\sin\theta_{13}$,
which typically implies $s^2_{13}\gtap$0.01. 
Consider two simple cases \cite{FPR}.\\
$\bullet$ Hierarchy of $\lambda_{ij}$:
$\lambda_{12}\equiv \lambda$, $\lambda_{23} \sim \lambda^{2}$, 
$\lambda_{13} \sim \lambda^{3}$.\\
Then $s_{13} \cong \lambda/\sqrt{2}$, 
$\sin^22\theta_{23}$=1-16$s^4_{13}$,
$\tan^2\theta_{12}$=1-4$s_{13}($1-2$s_{13}$+$4s^2_{13})$, and  
$\sin^22\theta_{23}\gtap$0.96; we also have
$s^2_{13}\gtap 0.01$ for $\tan^2\theta_{12}\leq 0.58$ \cite{BCGPRKL2}.\\
$\bullet$ $\lambda_{12} \equiv \lambda$,
$\lambda_{23} \sim \lambda/2$ (mild hierarchy), 
$\lambda_{13} \sim \lambda^{3}$.\\ 
Now $\sin^22\theta_{23}\gtap$0.90.\\ 
If $\lambda_{23}$=1, $\lambda_{12(13)}\ltap$0.35,
one has $\sin^22\theta_{23}\gtap$0.997, while for
$\lambda_{ij} \geq \frac{1}{\sqrt{2}}$, $\sin^22\theta_{23}\gtap$0.95
\cite{FPR}. Obviously, a sufficiently precise
measurement of $\sin^22\theta_{23}$ would allow
to distinguish between the three possibilities.

   If CP is not conserved, we have \cite{PPR3}

\vspace{-0.3cm}
$$\pmns = U_{\rm L}^\dagger U_\nu =  
\tilde{U}_{\rm lep}^\dagger P_\nu \tilde{U}_\nu Q_\nu,$$

\vspace{-0.1cm}
\noindent where, in general, 
$\tilde{U}_{\rm lep}$, $\tilde{U}_{\nu}$ are CKM-like
matrices each containing 3 angles and 1 $CPV$ phase, 
$P_{\nu}={\rm diag} (1,e^{i \phi},e^{i \omega})$ and
$Q_{\nu}={\rm diag} (1,e^{i \rho},e^{i \sigma})$.
Suppose that $\tilde{U}_\nu = U_{\rm bimax}$ and arises
from diagonalization of the simplest 
possible $M_{\nu}$ \cite{FPR},
\vspace{-0.2cm}
\be \label{eq:lelmlt}
M_\nu = \frac{m}{\sqrt{2}} \, 
\left( 
\bad 
0 & e^{-i\alpha'} & e^{-i\beta'} \\[0.2cm]
e^{-i\alpha'} & 0 & \epsilon~e^{-i\gamma'} \\[0.2cm]
 e^{-i \beta'} & \epsilon~e^{-i\gamma'} & 0  
\ea 
\right)~
\ee

%
\noindent Here $\alpha',\beta',\gamma'$ are phases,
$m^2 \cong -\dma$=$\Delta m^2_{23}$ and
$\dmsol$=$\Delta m^2_{21} \cong \sqrt{2}\epsilon m^2$,
$\epsilon \sim 0.03$.
The $\nu$-mass spectrum is $IH$.
In the limit $\epsilon = 0,\tilde{U}_{\rm lep} = {\bf 1}$,
$L' = L_e - L_{\mu} - L_{\tau}$ is conserved \cite{STP82PD}.
For $\tilde{U}_{\rm lep} \neq {\bf 1}$, 
$(\alpha' - \gamma')$=$\omega$ and $(\beta' - \gamma')$=$\phi$ 
are physical $CPV$ phases, $Q_{\nu} = {\bf 1}$, 
and $U_{\rm PMNS}$=$\tilde{U}_{\rm lep}^\dagger P_\nu U_{\rm bimax}$.
The Dirac and Majorana phases in $\pmns$ 
have the same ``source'' - the $CPV$ phases in 
$\tilde{U}_{\rm lep}^\dagger$ and $P_\nu$,
$\psi$ and $\phi$, $\omega$. Even $\theta_{ij}$
depend on the latter. 

 For, e.g., ``small'' 
$\lambda_{12}$=$\lambda$, $\lambda_{23}$=$ A\lambda$,
$\lambda_{13}$=$B\lambda^3$, $A,B\sim 1$, one finds
up to terms ${\cal{O}}(\lambda^3)$ \cite{FPR}:\\ 
$\tan^2 \theta_{12}\cong 1 - 2 \sqrt{2}\lambda c_\phi +  
2 (2 c^2_\phi - \sqrt{2}A c_\omega)\lambda^2$,\\
$\sqrt{2}s_{13}\cong \lambda$--$A\lambda^2c_{\omega - \phi}$,
$\sin^2 2\theta_{23}\cong$1--4$A^2\lambda^2c_{\omega - \phi}$,\\ 
%
where $c_\phi$=$\cos \phi$, etc.\\
The rephasing invariants associated with $CPV$ phases
$\delta$, $\beta$ and $\alpha$-$\beta$ read, respectively: 
$J_{CP} = 
{\rm Im} \left\{ U_{e1} \, U_{\mu 2} \, U_{e 2}^\ast \, U_{\mu 1}^\ast 
\right\}$ \cite{CJ85,PKSP3nu88},\\ 
$S_1$=${\rm Im}\left\{ U_{e1}U_{e3}^\ast \right\}$,
$S_2$=${\rm Im}\left\{ U_{e2}U_{e3}^\ast \right\}$ \cite{JMaj87}.\\ 
$J_{CP}$ controls the magnitude of 
$CP$ and $T$ violating effects in 
$\nu$-oscillations \cite{PKSP3nu88}; 
$S_1$, $S_2$ appear in the effective Majorana mass 
$\meff$ in \betabeta-decay \cite{BPP1}.
In general, $J_{CP}$, $S_1$ and $S_2$ are independent.
However, in the scheme with approximate
$L_e - L_{\mu} - L_{\tau}$ symmetry we are considering,
and to leading order in $\lambda$ we find \cite{FPR}:

\vspace{-0.3cm}
$$J_{CP} 
\cong \frac{S_1}{2\sqrt{2}} \cong \frac{S_2}{2\sqrt{2}}.$$

\vspace{-0.1cm}
\noindent Thus, {\it the magnitude of CP--violating effects in 
$\nu$-oscillations is directly related 
to the magnitude of CP--violating effects associated
with the Majorana nature of neutrinos}. 
One also finds \cite{FPR}:

\vspace{-0.4cm}
$$\meff \cong  \sqrt{|\Delta m^2_{\rm A}|}
~\left |\cos2\theta_{\odot} + i~8J_{CP} \right |~,$$ 

\vspace{-0.2cm}
\noindent 
i.e., $J_{CP}$ determines the deviation 
of $\meff$ from its minimal value 
(for $IH$ spectrum) \cite{BPP1}.

  The approach to understand the pattern of 
$\nu$-mixing discussed above is by no means unique 
(see, e.g., \cite{FFerug,CAlbr}). 
It demonstrates that Dirac and Majorana
$CPV$ phases (and effects) 
can be related, and that 
$\theta_{ij}$ can depend on $CPV$ phases.

\vspace{-0.30cm}
\section{Comments on Future Progress}

\vspace{-0.20cm}
 Future progress in the studies of  
$\nu$-mixing will be crucial
for understanding at fundamental level
the mechanism generating it. 
The requisite data is forseen
to be provided by\\
$\bullet$ $\nu_{\odot}-$ and $\nu_{\rm A}-$ experiments:\\
SK, SNO, SAGE, BOREXINO \cite{BOREXINO}, LowNu 
\cite{LowNu}, and 
SK ($\nu_{\rm A}$), MINOS ($\nu_{\rm A}$) \cite{MINOS}, 
INO ($\nu_{\rm A}$) \cite{INO};\\
$\bullet$ Reactor $\bar{\nu}_e$ experiments with $L\sim$(1 - 180) km;\\
$\bullet$ Accelerator experiments:\\
K2K ($L\sim$250 km), MINOS ($L\sim$730 km),
OPERA and ICARUS \cite{OPERA,ICARUS} ($L\sim$730 km);\\ 
$\bullet$ Experiments with super beams:\\
T2K ($L\sim$295 km) \cite{T2K},
NO$\nu$A ($L\sim$800 km) \cite{NOnuA}\\
SPL+$\beta-$beams with UNO (1 megaton) detector
(CERN-Frejus, $L\sim$135 km) \cite{SPLbeta};\\
$\bullet$ $\nu$-Factory experiments ($L\sim$3000;7000 km) \cite{NuFact};\\
$\bullet$ $(\beta\beta)_{0\nu}-$
and $^{3}$H $\beta-$ decay experiments \cite{bb0nu,H3};\\
$\bullet$ Astrophysical/cosmological 
observations \cite{Astro}.\\ 

\vspace{-0.3cm}
\noindent \underline{\bf Absolute Neutrino Mass Measurements.} 
The Troitzk and Mainz \hbeta experiments 
provided information on the
$\bar{\nu}_e$ mass \cite{H3}:
$m_{\bar{\nu}_e}<$2.2 eV at 95\%~C.L.
The KATRIN experiment \cite{H3}
expected to start in 2007,
is planned to reach sensitivity to 
$m_{\bar{\nu}_e}\sim$0.20 eV (95\% C.L.),
and thus to probe the region of 
QD $\nu$-mass spectrum.  
In this region $m_{1,2,3} \cong m_0 \cong m_{\bar{\nu}_e}$.

  The CMB data of the WMAP experiment 
were used to obtain an upper limit 
on $\sum_{j} m_{j}$ \cite{Astro}:\\ 
$\sum_{j} m_{j} <$ (0.7--1.8) eV   (95\%~C.L.).\\
The WMAP and future PLANCK 
experiments can be sensitive to 
$\sum_{j} m_{j} \cong 0.40$ eV.
Data on weak lensing of galaxies 
by large scale structure,
combined with WMAP and PLANCK data,
may allow one to determine 
$\sum_{j} m_{j}$ with an uncertainty 
of $\sim$(0.04--0.10) eV (see, e.g., \cite{Astro}).\\

\vspace{-0.30cm}
\noindent \underline{\bf $\betabeta-$Decay Experiments}.
The $\betabeta$-decay experiments \cite{bb0nu}
have a remarkable physics potential
\footnote{For detailed discussion and list
of references see, e.g., \cite{STPFocusNu04}}.
They can establish the Majorana nature of 
neutrinos $\nu_j$. If $\nu_j$
are Majorana particles,
they can provide unique
information \\
$\bullet$ on the type of
$\nu$-mass spectrum \cite{BGGKP99,BPP1,PPSNO2bb},\\  
$\bullet$ on the absolute scale of $\nu$-masses \cite{PPW}, and\\
$\bullet$ on the
Majorana $CPV$ phases in $\pmns$ \cite{BGKP96,PPR1}. 
The $\betabeta$-decay, 
(A,Z)$\rightarrow$(A,Z+2)+$e^-$+$ e^-$, 
is allowed if 
$\nu_j$ are Majorana particles 
(see, e.g., \cite{BiPet87}).
The nature - Dirac or Majorana, of 
neutrinos $\nu_j$ is related 
to the fundamental symmetries of 
particle interactions. Neutrinos 
$\nu_j$ will be Dirac fermions if 
particle interactions conserve 
the total lepton charge $L$. 
They can be Majorana particles
if there does not exist 
any conserved lepton charge.
Massive neutrinos are 
predicted to be of Majorana nature
by the see-saw mechanism of 
$\nu$-mass generation 
(see, e.g., \cite{seesaw}),
which also provides 
an attractive explanation of the
smallness of 
$\nu$-masses and 
- through leptogenesis theory 
(see \cite{LeptoG}), of the 
baryon asymmetry 
of the Universe.

 If $\nu_j$ are Majorana fermions,
the $\betabeta$-decay amplitude  
of interest has the form 
(see, e.g., \cite{BPP1}):\\
$A\betabeta \cong \mefff~M$, where 
$M$ is the corresponding 
nuclear matrix element (NME) and

\vspace{-0.55cm}
$$\meff=\left| m_1 |U_{\mathrm{e} 1}|^2 
+ m_2 |U_{\mathrm{e} 2}|^2e^{i\alpha}
 + m_3 |U_{\mathrm{e} 3}|^2e^{i\beta} \right|$$

\vspace{-0.1cm}
\noindent $|U_{\mathrm{e}1}|$=$c_{12}c_{13}$,
$|U_{\mathrm{e}2}|$=$s_{12}c_{13}$, 
$|U_{\mathrm{e}3}|$=$s_{13}$.
In the case of $CP$-invariance one has \cite{LW81},\\  
$\eta_{21} \equiv e^{i\alpha}$=$\pm 1$, 
$\eta_{31}\equiv e^{i\beta}$=$\pm 1$, $\eta_{21(31)}$ being the \\ 
\noindent relative 
CP-parity of Majorana neutrinos 
$\nu_{2(3)}$ and $\nu_1$. 
Thus, $\meff$ depends \cite{SPAS94}
on $\theta_{\odot}$, $\theta_{13}$,
$\dma$, $\dmsol$ as well as on 
$min(m_j)$, Majorana 
phases $\alpha$, $\beta$ and the $\nu$-mass spectrum.
The predicted value of $\meff$ for 
$\sin^2\theta_{13}$=0.04 and 
90\% C.L. allowed 
values of $\dma$ and 
$\dmsol$, $\theta_{\odot}$ (Fig. 1)
is shown as function of
$min(m_j)$ in Fig. 2.
\begin{figure}[htb]
\vskip -0.5cm
\includegraphics[width=7cm,height=5cm,clip=]{KL2meff90SK3nus1304.epsi}
\vskip -0.5cm
\caption{The value of $\meff$ 
as function of $min(m_j)$
for $\sin^2\theta_{13}$=0.04 and 
90\% C.L. allowed ranges of values
of $\dma$, $\dmsol$ and 
$\theta_{\odot}$ \cite{PPSNO2bb,SPNobSymp04}.
}
\label{Fig2}
\end{figure}

  The main features of the predictions for $\meff$ 
are \cite{PPSNO2bb,PPW,STPFocusNu04}: i) for $NH$ spectrum, 
typically $\meff\ltap$0.006 eV, and
$\meff\sim$0 is possible;
ii) for $IH$ spectrum, $\meff\gtap\sqrt{|\dma|}\cos2\theta_{\odot}
\gtap$0.012 eV and $\meff\ltap\sqrt{|\dma|}\ltap$0.06 eV; 
iii) in the case $QD$ spectrum, 
$\meff\gtap m_0\cos2\theta_{\odot}\gtap$0.05 eV 
and $\meff\ltap m_0$, with $m_0\gtap$0.2 eV and
$m_0<$2.2 eV \cite{H3}, $m_0<$0.6 eV \cite{Astro}. 
Thus, for $IH$ and $QD$ spectra, $\meff$
is limited from below. 

  Many experiments have searched for
$\betabeta$-decay \cite{Morales02}. 
The best sensitivity was achieved in\\ 
Heidelberg-Moscow $^{76}$Ge experiment 
\cite{HMGe76}. A positive signal at $>$3$\sigma$,
corresponding to\\  
$\meff = (0.1 - 0.9)~{\rm eV}$  at 99.73\% C.L.,
is claimed to be observed \cite{HMGe76}. 
Two experiments, NEMO3 (with $^{100}$Mo and 
$^{82}$Se) \cite{NEMO3}
and CUORICINO (with $^{130}$Te) \cite{CUORI},
designed to reach sensitivity to 
$\meff\sim$(0.2-0.3) eV, announced 
first results:
$\meff<$ (0.7--1.2) eV \cite{NEMO3} and $\meff<$ (0.3 -- 1.6) eV
\cite{CUORI} (90\% C.L.),
where estimated uncertainties in the $NME$
\cite{bb0nuNME} are accounted for.
A number of projects aim at sensitivity to 
$\meff\sim$(0.01--0.05) eV \cite{bb0nu}:
CUORE ($^{130}$Te), GENIUS ($^{76}$Ge),
EXO ($^{136}$Xe), MAJORANA ($^{76}$Ge),
MOON ($^{100}$Mo), 
XMASS ($^{136}$Xe), etc. These experiments
will probe the region corresponding
to $IH$ and $QD$ spectra
and test the positive result 
claimed in \cite{HMGe76}. 
The knowledge of the relevant $NME$
with sufficiently small uncertainty
is crucial for obtaining 
quantitative information on
the $\nu$-mixing parameters 
from a measurement of
$\betabeta$-decay half-life.
{\it In view of their importance for
understanding the origin of
$\nu$- masses and mixing,
 performing few 
$\betabeta$-decay 
experiments with sensitivity to
$\meff\sim$(0.01--0.05) eV 
(or better)
should have highest priority in the 
future studies of $\nu$-mixing.}\\

\vspace{-0.30cm}
\noindent \underline{\bf The CHOOZ Angle $\theta_{13}$}.
The angle $\theta_{13}$ plays extremely
important role in the phenomenology of 
$\nu$-oscillations. It controls together 
with $\sin\delta$ the magnitude of
$CP$- and $T$-violating effects 
in $\nu-$oscillations.
It controls the sub-dominant 
$\nu_{\mu} \leftrightarrow\nu_e$ and
$\bar{\nu}_{\mu} \leftrightarrow \bar{\nu}_e$ 
oscillations of $\nu_{\rm A}$
and in the accelerator experiments 
MINOS, OPERA, ICARUS, T2K, NO$\nu$A, etc. 
and at $\nu-$factories.
The value of $\meff$ in $\betabeta$-decay
in the case of $NH$ spectrum
depends on $\sin^2\theta_{13}$ \cite{BPP1,PPW}.
The knowledge of $\theta_{13}$ 
is crucial for finding the correct theory 
of $\nu$-mixing as well.

  If $\sin^22\theta_{13} \ltap (0.01-0.008)$,
the $CPV$ effects in $\nu$-oscillations
would be too small to be observed in
T2K and NO$\nu$A experiments 
\cite{TMU04,MShae04}.
Thus, the future program
of searches for $CPV$ effects
in accelerator experiments
depends critically on 
the value of $\theta_{13}$.
The sensitivity of future experiments with
{\it conventional beams (MINOS+ICARUS+OPERA)},
with off-axis super beams, T2K-SK, NO$\nu$A (NuMI),
and with reactor $\bar{\nu}_e$ - Double-CHOOZ 
\cite{Reactor04},
to $\sin^22\theta_{13}$, as function of 
$|\Delta m^2_{31}|$, is illustrated in Fig. 3.
\begin{figure}[htb]
\vskip -0.5cm
\includegraphics[width=7.5cm,height=6cm,clip=]{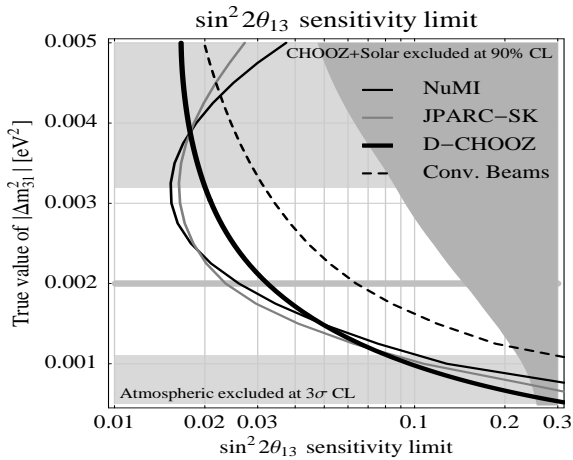}
\vskip -0.5cm
\caption{The sensitivity of 
future experiments to $\sin^22\theta_{13}$
\cite{TMU04}.
}
\label{Fig3}
\end{figure}

\vspace{-0.35cm}
 There are several proposals for
reactor $\bar{\nu}_e$ experiments 
with baseline $L\sim$(1-2) km 
\cite{Reacth13,Reacth13J}, 
which could imporve the current limit 
$\sin^2\theta_{13}<$0.05 by a 
factor of (5-10) \cite{Reactor04}. 
The most advanced in preparation 
is the Double-CHOOZ project. 
{\it The reactor $\theta_{13}$ 
experiments, in our view,
should have highest priority
in the program 
of research in $\nu$-physics:
they can compete in sensitivity with
accelerator experiments (T2K-SK, NO$\nu$A) and 
can be done on relatively short 
(for experiments in this field) 
time scale. The planning of 
experiments to study $CPV$ effects 
in $\nu$-oscillations 
would benefit 
significantly from the results
of a high precision reactor 
$\theta_{13}$ experiment.} \\

\vspace{-0.30cm} 
\noindent \underline{\bf Measuring $\dmsol \equiv \Delta m^2_{21}$ and 
$\theta_{\odot} \equiv \theta_{12}$}.
The current solar 
neutrino and KL 766.3 Ty 
data determine $\Delta m^2_{21}$
and $\sin^2\theta_{12}$ 
with uncertainties of 12\% and 24\% at 3$\sigma$ 
\cite{BCGPTH1204}.
Accounting for possible 
reduction of 
errors in the data 
from the phase-III of 
SNO experiment \cite{SNO123}
could lead to a reduction 
only of the error in $\sin^2\theta_{12}$ 
to 21\% \cite{SKGdCP04,BCGPTH1204}. 
If instead of 766.3 Ty
one uses simulated 3 kTy
KL data in the same 
data analysis, the 3$\sigma$ errors in 
$\Delta m^2_{21}$ and $\sin^2\theta_{12}$ 
diminish to 7\% and 18\% \cite{BCGPTH1204}. 

  The most precise measurement of 
$\Delta m^2_{21}$ could be achieved 
\cite{SKGdCP04} using SK
doped with 0.1\% of Gadolinium 
(SK-Gd) for detection of reactor 
$\bar{\nu}_e$ \cite{SKGdBV04}:
SK gets the same flux of reactor
$\bar{\nu}_e$ as KamLAND and
after 3 years of 
data-taking, $\Delta m^2_{21}$
would be determined with a 
3.5\% error at 3$\sigma$ 
\cite{SKGdCP04}. 
A dedicated reactor  
$\bar{\nu}_e$ experiment with a 
baseline $L\sim$60 km tuned to 
$\bar{\nu}_e$ survival 
probability minimum, 
could provide the most precise 
determination of $\sin^2\theta_{12}$ 
\cite{TH12,HMinth1204,BCGPTH1204}:
with statistics of $\sim$60 GWkTy 
and systematic error 
of 2\% (5\%), $\sin^2\theta_{12}$  
could be measured with uncertainty
of 6\% (9\%) at 3$\sigma$ \cite{BCGPTH1204}. 
A generic LowNu $\nu-e$ elastic
scattering experiment, 
designed to measure the $pp$ 
$\nu_{\odot}$-flux with an 
error of 3\% (1\%), would permit  
to determine $\sin^2\theta_{12}$ 
with an error of 14\% (19\%) at 3$\sigma$ 
\cite{BCGPTH1204}. 
The inclusion of the uncertainty
in $\theta_{13}$
($\sin^2\theta_{13}<$0.05)
in the analyses increases the 
quoted errors by (1-3)\% 
\cite{BCGPTH1204}.\\

\vspace{-0.30cm} 
\noindent \underline{\bf Measuring $|\dma|$, 
$\theta_{\rm A}\equiv\theta_{23}$,
$sign(\dma)$ and $\delta$}.
The expected 3$\sigma$ uncertainties in 
$|\dma|$=$|\Delta m^2_{31}|$
from studies
of $\nu_{\mu}$-oscillations in
i) MINOS + CNGS \cite{MINOS,OPERA,ICARUS},
ii) NO$\nu$A \cite{NOnuA} and
iii) T2K (SK) \cite{T2K} experiments,
if the true $|\Delta m^2_{31}|$=
2$\times 10^{-3}$ eV$^2$ and
true $\sin^2\theta_{23}$=0.5, 
read \cite{TMU04}: 
i) 26\%, ii) 25\% and iii) 12\%.
T2K (SK) and NO$\nu$A experiments
will measure also $\sin^22\theta_{23}$
with a high precision - (1-2)\% at 1$\sigma$.
However, they would not be able to
resolve the $\theta_{23}$--$(\pi/2 - \theta_{23})$
ambiguity if $\sin^22\theta_{23}<$1.
T2K and NO$\nu$A are planned to begin 
in 2009 and 2011 (or 2009 \cite{NOnuA}) and
in their first phase, both 
experiments will use $\nu_{\mu}$-beams.  
The data from this phase will not allow
to determine the $sign(\Delta m^2_{31})$.
Even if $\sin^22\theta_{13}\sim$0.10,
without the knowledge of
$sign(\Delta m^2_{31})$
it would be impossible 
to get unambiguous information
on $CP$-violation in 
$\nu$-oscillations (induced by
$\delta$) using only the phase-I 
T2K and NO$\nu$A data 
\cite{TMU04}. If $\sin^22\theta_{13}\gtap$0.05,
information on $sign(\Delta m^2_{31})$
and $\sin^2\theta_{23}\gtap$0.5
might be obtained in $\nu_{\rm A}$-experiments
by studying the Zenith angle
dependence of the multi-GeV 
$e$- and $\mu$- like, and/or
$\mu^{\pm}$, events \cite{JBSP203}.
Resolving all parameter degeneracies
\cite{PDeg,MShae04} and determining 
whether $\delta$ takes a $CPV$ value if 
$\sin^22\theta_{23}<$1 and no
information on $sign(\Delta m^2_{31})$
is available, would be a 
formidable task and would 
require high statistics 
(phase-II) data on $\nu_{\mu}$- 
and $\bar{\nu}_{\mu}$-
oscillations both from T2K and
NO$\nu$A and data on $\theta_{13}$ 
from a reactor experiment 
\cite{TMU04,PDeg,MShae04};
if $\sin^22\theta_{13}\ltap$0.01, 
data from SPL+$\beta$-beam 
experiments \cite{SPLbeta} 
or from experiments at 
$\nu$-factory \cite{NuFact}
might be required.

\vspace{-0.20cm}
\section{Instead of Conclusions}

\vspace{-0.2cm}
We are at the beginning of the road 
leading to a 
comprehensive understanding 
of the patterns of 
 \begin{figure}[htb]
\includegraphics[width=8cm,height=6.5cm,clip=]{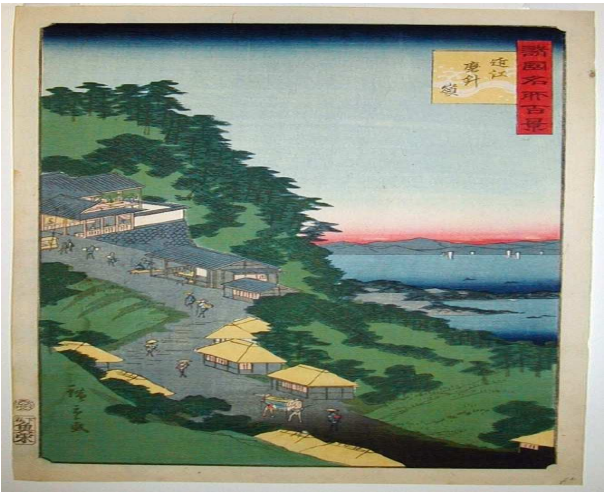}   
\caption{Landscape by Y. Hiroshige (Ukiyoe master, Edo epoch, Japan).}
\label{Fig4} 
\end{figure}

\vspace{-0.4cm}
\noindent neutrino masses and mixing and 
of their origin. 
The road is not easy and 
we do not quite know how long
our ``journey'' will take, 
how difficult it will be 
and what we will finally discover.  
However, I am sure the 
``view'' that will open to us
from the ``summit'' at 
the end of this ``journey'' 
will be of dazzling clarity,
perspective and beauty (Fig. 4).\\

\vspace{-0.30cm}
{\bf{Acknowledgements}}. It is a pleasure 
to thank the Organizers of the Neutrino 
2004 International Conference for 
assembling such a scientifically 
enjoyable meeting.


\end{document}